\documentclass[aps,prl,twocolumn,showpacs,amssymb, footinbib]{revtex4-1}
\usepackage{amsfonts,amsmath,amssymb,amsthm,bbm,hyperref}
 
\usepackage{color,psfrag}
\usepackage[dvips]{graphicx}

\usepackage[normalem]{ulem}

\newcommand{\m}{\mbox{}}

\newcommand{\be}{\begin{equation}}
\newcommand{\ee}{\end{equation}}
\newcommand{\ba}{\begin{eqnarray}}
\newcommand{\ea}{\end{eqnarray}}

\begin{document}

\title{Black hole entropy from loop quantum gravity in higher dimensions}

\author{N. Bodendorfer}
\email[]{norbert@gravity.psu.edu} 

\affiliation{Institute for
Gravitation and the Cosmos \& Physics Department, Penn State, University Park, PA 16802, U.S.A.}

\date{{\small July 19, 2013}}

\begin{abstract}

We propose a derivation for computing black hole entropy for spherical non-rotating isolated horizons from loop quantum gravity in four and higher dimensions. The state counting problem effectively reduces to the well studied $3+1$-dimensional one based on an SU$(2)$-Chern-Simons theory, differing only in the precise form of the area spectrum. 

\end{abstract}

\pacs{04.50.Gh, 04.60.-m, 04.70.Dy }

\maketitle

\section{Introduction}

The calculation of black hole entropy is one of the main tests for candidate theories of quantum gravity. Different gravitational theories, horizon topologies and higher dimensions offer a great variety of non-trivial tests for such candidate theories. Despite this plethora of examples, the resulting entropies, calculated by different approaches, are very similar. At the classical level, this is explained by the Wald formula  \cite{WaldBlackHoleEntropy}. Carlip \cite{CarlipBlackHoleEntropyFromConformalFieldTheoryInAnyDimension} and Solodukhin \cite{SolodukhinConformalDescriptionOf} noticed that one can derive black hole entropy in any dimension using conformal field theory techniques. This derivation also results in the universal logarithmic corrections $-3/2 \, \text{Log} (A)$ \cite{KaulLogarithmicCorrectionTo, CarlipLogarithmicCorrectionsTo}. Still, it would be desirable to have a derivation from a fundamental quantum theory of gravity.

Loop quantum gravity (LQG) \cite{RovelliQuantumGravity, ThiemannModernCanonicalQuantum} has matured into a serious candidate for such a theory.
Here, the calculation of black hole entropy is translated to the calculation of the entropy of isolated horizons, see \cite{Diaz-PoloIsolatedHorizonsAnd} for a review. The advantage of this viewpoint is that isolated horizons provide local definitions of the relevant horizons (e.g. black hole or cosmological) and an explicit knowledge of their complete past and future is not required.

The main non-trivial point which allows to compute the entropy of black holes within LQG is that the connection variables used induce new degrees of freedom which are associated to the isolated horizon and not the to bulk. 
The classical part of this derivation was extended to higher dimensions in \cite{BTTXII} and to Lanczos-Lovelock gravity in \cite{BNII}, where the recently introduced connection variables for higher-dimensional general relativity \cite{BTTI, BTTII, BTTIII, BTTIV} were employed. While the horizon degrees of freedom can be rewritten in a form similar to a higher-dimensional Chern-Simons theory, it turns out to be more economical to use a canonically conjugate pair of normals $n^I$ and $\tilde s^I$ as horizon degrees of freedom \cite{BTTXII}.
In this paper, we will use the latter formulation in order to generalize the entropy calculation to higher dimensions. It turns out that the problem of computing the entropy reduces largely to the one familiar from $3+1$ dimensions due to the fact that the horizon Hilbert spaces in different dimensions can be mapped 1 to 1 onto each other.

\section{Isolated horizon degrees of freedom}		

	\label{sec:IHDOF}

In this section, we will recall the isolated horizon degrees of freedom using the dimension-independent connection variables as derived in \cite{BTTXII, BNII}. 
We will neglect the detailed treatment of spatial infinity in this paper, as it is not relevant for the discussion, see e.g. \cite{AshtekarIsolatedHorizonsThe}. 
The main result is that one can perform a phase space extension from the Lorentzian ADM phase space \cite{ArnowittTheDynamicsOf} in $D+1$ spacetime dimensions to a canonical pair consisting of an SO$(D+1)$ connection $A_{aIJ}$ and its conjugate momentum $\pi^{aIJ}$. Here, $a,b = 1,\ldots,D$ are tensor indices on a spatial slice $\Sigma$ and $I,J=0,\ldots,D$ are internal SO$(D+1)$ indices. Let $q_{ab}$ be the induced metric on $\Sigma$ and $K_{ab}$ the extrinsic curvature, which together parametrize the ADM phase space. We introduce the $(D+1)$-bein $e_{a}^I$ satisfying $q_{ab} = e_{a}^I e_{b}^J \delta_{IJ}$ and the hybrid spin connection $\Gamma^\text{H}_{aIJ}$ annihilating it \cite{PeldanActionsForGravity}. 
The information of $q_{ab}$ and $K_{ab}$ is contained in the connection variables as 
\begin{eqnarray}
	(A_{aIJ} - \Gamma^\text{H}_{aIJ}) \pi^{bIJ}& \approx & 2 \sqrt{q} q^{bc} K_{ac} \\
	\pi^{aIJ} & \approx & 2 / \beta \, n^{[I} e_{b}^{J]} q^{ab} \sqrt{q} \text{,}
\end{eqnarray}
where $\beta \in \mathbb{R} \backslash  \{0\}$ is a free constant. $\approx$ means equality on the constraint surface, defined by the SO$(D+1)$ Gau{\ss} law 
and the simplicity constraint $\pi^{a[IJ} \pi^{b|KL]} = 0$. 
The simplicity constraint ensures that $\pi^{aIJ} \approx 2 / \beta \, n^{[I} e_{b}^{J]} q^{ab} \sqrt{q}$, that is, it is the product of a normal $n^I$ and a densitized $D+1$-bein orthogonal to $n^I$, see \cite{FreidelBFDescriptionOf, BTTI, BTTII} for further details, e.g. a topological sector in $3+1$ dimensions.

One can check that the Poisson brackets
\begin{alignat}{4} \label{eq:NewPoisson}
	\{A_{aIJ}(x), \pi^{bKL}(y) \} &= \delta_{a}^b \delta^{KL}_{IJ} \delta^{(D)}(x-y) ~~~&& \text{on $\Sigma$} 
\end{alignat}
reproduce the ADM Poisson brackets on the constraint surface, thus confirming equivalence to the ADM formulation. The construction works analogously for SO$(1,D)$ as gauge group, however the compactness of SO$(D+1)$ is preferred for quantization purposes.

In the presence of a boundary $H$ of $\Sigma$ (here, the intersection of $\Sigma$ with an isolated horizon $\Delta$) with boundary unit normal $s^a$ pointing outward of $\Sigma$, the above calculation is only correct up to a boundary term, which results in the additional Poisson bracket
\be
	\{\tilde{s}^I(x), n_J(y)  \} = \beta  \, \delta^{I}_{J} \delta^{(D-1)}(x-y) ~~~ \text{on $H$}	 \label{eq:BoundaryPoisson}
\ee
with $\tilde{s}_J := \sqrt{h} s^a e_{aJ}$ and $h = \det h_{\alpha \beta}$, $\alpha, \beta = 1,\ldots,D-1$ being the determinant of the induced metric $h_{\alpha \beta}$ on $H$ \footnote{It should be noted that in order for $n^I$ to be a unit vector, the Poisson bracket should be modified slightly to ensure consistency with $n^I n_I = 1$. This already follows from reconstructing this bracket from \eqref{eq:SympStrucL}. We will comment on this detail in a future publication.}. Classically, the product $2 n^{[I} \tilde s^{J]}$ is determined by continuity from the bulk fields as 
\be
	2 / \beta \, n^{[I} \tilde s^{J]} = \hat s_a \pi^{aIJ} \text{,} \label{eq:BoundaryCondition}
\ee
where $\hat s_a = s^b q_{ab} \sqrt{h/q}$ is an appropriately densitized co-normal. \eqref{eq:BoundaryCondition} is the analogue of the isolated horizon boundary condition $F^{IJ}(A) \propto e^I \wedge e^J$ familiar from the $3+1$-dimensional treatment \cite{Diaz-PoloIsolatedHorizonsAnd}. In fact, \eqref{eq:BoundaryCondition} can be rewritten in the form $F^{IJ}(A) \propto e^I \wedge e^J$ in $3+1$ dimensions 
\cite{BTTXII}.

Since $n^I \tilde{s}_I = 0$ classically by construction, the information contained in these variables associated to the boundary is the $(D-1)$-area-density $\sqrt h = \sqrt{\tilde s^I \tilde s_I}$ of $H$. 
We consider the densitized bi-normals $L^{IJ} := 2 / \beta \, n^{[I} \tilde s^{J]}$ since they already contain this information. Their Poisson algebra is given by 
\begin{eqnarray}
	\{ L_{IJ}(x), L_{KL}(y) \}  = 4 \,   \delta^{(D-1)}(x-y) \delta_{L][I} L_{J][K}  \label{eq:SympStrucL} 
\end{eqnarray}
and thus by the Lie algebra so$(D+1)$ at every point of $H$. We will base our quantization of the horizon degrees of freedom on this algebra. 

A similar structure has been found for Ashtekar-Barbero variables in \cite{EngleBlackHoleEntropyFrom}, where it was also emphasized that compatibility of \eqref{eq:NewPoisson}, \eqref{eq:BoundaryCondition}, and \eqref{eq:SympStrucL} leads us to use holonomy-flux variables, thus strengthening the uniqueness result \cite{LewandowskiUniquenessOfDiffeomorphism, FleischhackRepresentationsOfThe} of the Ashtekar-Isham-Lewandowski representation based on these variables even further.

\section{Conditions on $\Delta$}

We demand $\Delta$ to be a spherical non-rotating isolated horizon, see \cite{BTTXII} for details. From this, it follows that the null normal $l^\mu$, $\mu = 0,\ldots,D$, to the isolated horizon $\Delta$ satisfies $\nabla_\alpha l^\mu = \omega_\alpha l^\mu=  0$, i.e. $l^\mu$ is covariantly constant on $H$. 
For the SO$(D+1)$-based connection variables used in this paper, we find $\nabla_\alpha k^I = 0$, with $k^I = (n^I + s^I)/\sqrt{2}$ and $s^I = s^a e_a^I$, where the connection used is the pullback of the (bulk) SO$(D+1)$-connection on $\Sigma$ to $H$. 
Note that the internal vectors $k^I$ and $l^I = (n^I - s^I)/\sqrt{2}$ cannot be null, i.e. $k^2 = l^2 = 1$, $l \cdot k = 0$, and are thus not related to the spacetime null-vectors $l^\mu, k_\mu$ \footnote{For SO$(1,D)$ as internal gauge group, we would have $ \nabla_{\alpha} l^I = 0$ with $l^2 = 0$.}.
Since $n^{[I} \tilde s^{J]} = - k^{[I} \tilde l^{J]}$, we demand that the bi-normals $L^{IJ}$ share a common vector, here $k^I$, to incorporate this classical property of an isolated horizon. 
We can cover $H$ with two contractible charts due to its spherical topology. 
On each of these charts, the gauge $k^I$ = const. is accessible. Moreover, $A_\alpha \m^I \m_J k^J = 0$ in this gauge, so that there is a  trivial identification of the $k^I(x)$ at different points of the two charts. 
While a more elaborate treatment might be desirable (see the discussion section), this leads us to demand the ``off-diagonal'' horizon simplicity constraints $L_{[IJ}(x) L_{KL]}(y) = 0$ for $x$, $y$ in the same chart, which should restrict us to the relevant degrees of freedom.
We now need to investigate the constraint algebra.

\section{Constraint algebra}

The Gau{\ss} and spatial diffeomorphism constraints 
\begin{eqnarray}
	G^{IJ}[\Lambda_{IJ}] &=& - \int_{\Sigma} \pi^{aIJ} D_{a} \Lambda_{IJ} \label{eq:GaussLaw} + \int_H \Lambda_{IJ} L^{IJ} \\
	\mathcal{H}_a[N^a] &=& \frac{1}{2} \int_{\Sigma} \pi^{aIJ} \mathcal{L}_N A_{aIJ} + \frac{1}{\beta} \int_H  n^I \mathcal{L}_N \tilde s_I
\end{eqnarray}
form a closing algebra, where $\mathcal{L}_N$ denotes the Lie derivative along the shift vector $N^a$, satisfying $N^a \hat s_a = 0$ on $H$ to respect the boundary. Their associated vector fields are degenerate directions of the symplectic structure and they are consistent with \eqref{eq:BoundaryCondition}. The same is true for the simplicity constraint and the Hamiltonian constraint, provided that 
the lapse function vanishes on $H$. Thus, the Hamiltonian constraint does not have to be taken into account for the horizon Hilbert space, as in \cite{AshtekarIsolatedHorizonsThe}. 
Moreover, $A_H := \int_H \sqrt{h} \, d^{D-1}x$ is a Dirac observable.
As usual, see e.g \cite{AshtekarQuantumGeometryOf}, we assume that there exists at least one solution of the Hamiltonian constraint in the bulk which is compatible with a given configuration of punctures. Otherwise, a gauge fixing as in \cite{BSTI} could be employed. Up to here, the constraint algebra is closing (first class) \footnote{We refrain from restricting the gauge invariance on $H$, as e.g. done in \cite{AshtekarIsolatedHorizonsThe}, since the action principle detailed in \cite{BTTXII} does not require any such gauge fixing. This mimics as closely as possible the more physical case of spin networks extending inside the horizon \cite{KrasnovBlackHolesIn}.}. 

The horizon simplicity constraints need a more thorough investigation. They do not form a closing algebra, see e.g. \cite{BTTIII}, and are thus second class in Dirac's terminology \cite{DiracLecturesOnQuantum}. In order to be able to quantize them, we will select a maximally commuting subset adapted to the puncturing spin network in the next section, that is, we perform gauge unfixing \cite{MitraGaugeInvariantReformulationAnomalous}. 
This subset does however not form a closing algebra with the Gau{\ss} constraints, since the smearing functions $\Lambda_{IJ}$ are not constant on $H$. We thus further restrict to constant $\Lambda_{IJ}$ on $H$ and note that also in the U$(1)$ framework in $3+1$ dimensions \cite{AshtekarIsolatedHorizonsThe, AshtekarQuantumGeometryOf}, the restriction to constant $\Lambda_{IJ}$ on $H$ becomes necessary, however for different reasons \cite{ThiemannModernCanonicalQuantum}.

\section{Quantization}

While the $L_{IJ}$ are classically determined by the bulk fields through continuity, i.e. by the boundary condition \eqref{eq:BoundaryCondition}, this ceases to be the case in the quantum theory since the Hilbert space used is distributional by nature. Thus, in the same way as in $3+1$ dimensions, the $L_{IJ}$ have to be promoted to independent degrees of freedom in the quantum theory. 
The complete Hilbert space is therefore a product of the bulk Hilbert space defined in \cite{BTTIII} and the horizon Hilbert space, which is a product of SO$(D+1)$ representation spaces, on which the Poisson algebra \eqref{eq:SympStrucL} of boundary degrees of freedom can be represented as the generators of the gauge group SO$(D+1)$ \footnote{More precisely, one would regularize the $L^{IJ}$ like fluxes by smearing them over small surfaces on $H$.}.

The boundary condition \eqref{eq:BoundaryCondition} now has to be implemented as an operator. 
We construct fluxes by smearing $\hat s_a \pi^{aIJ}$ over a $(D-1)$-surface tangent to $H$. The fluxes can be promoted to quantum operators in the usual way \cite{BTTIII} and obey the same algebra as the $L_{IJ}$, which confirms the consistency of the framework. The quantum analog of the boundary condition \eqref{eq:BoundaryCondition} is automatically solved due to this algebraic structure by contracting holonomies ending on $H$ with a respective representation index of the horizon Hilbert space.
At points of $H$ where no spin network edge punctures, the boundary condition implies that the SO$(D+1)$ representation is trivial. The non-trivial part of the horizon Hilbert space is thus generated at the finite number of points where the bulk spin network punctures $H$.

The quantum Gau{\ss} law is also automatically solved in this way on $H$. We refer to \cite{ThiemannModernCanonicalQuantum} for an explicit quantization of the first term in \eqref{eq:GaussLaw}. Quantization of the second term cancels the first one.
Still, comparison with the $3+1$ dimensional treatment \cite{AshtekarQuantumGeometryOf} suggests that the horizon Hilbert space should be further restricted by a global gauge invariance condition, which is derived there by ``shrinking a loop around the back of the sphere''. 
This procedure is not readily available in higher dimensions. Instead, we follow the ideas of \cite{KrasnovBlackHolesIn} and enforce global gauge invariance in the next section by tracing over gauge related bulk degrees of freedom. 
Alternatively, we could quantize Gau{\ss}-invariant spin networks directly, which would also result in SO$(D+1)$ intertwiners as horizon states \footnote{We underline the discrepancy between solving the Gau{\ss} law at the classical and quantum level on $H$. In our context, we consider the classical solution to be the physical one, see also \cite{KrasnovBlackHolesIn}. A similar discrepancy is familiar from the simplicity constraint.}. This procedure however has the technical problem that the simplicity constraint is not gauge invariant, but only covariant, and still needs to be implemented in the quantum theory.

The spatial diffeomorphism constraint selects diffeomorphism equivalence classes of spin networks, see \cite{ThiemannModernCanonicalQuantum} for details.
The bulk simplicity constraint together with the boundary condition \eqref{eq:BoundaryCondition} restricts us to simple SO$(D+1)$ representations on $H$, which are labelled by a single non-negative integer $\lambda$ \cite{FreidelBFDescriptionOf, BTTIII}. This is consistent with the fact that we could have imposed diagonal horizon simplicity constraints on the horizon degrees of freedom.

We still have to deal with the second class nature of the off-diagonal horizon simplicity constraints. For this, we choose a maximally commuting subset by imagining the edges puncturing $H$ in each of the two charts to be connected to a spin network vertex and following the recipe in \cite{BTTV}.
It is easy to see that the subset is identical to a subset that would have been obtained from imposing the off-diagonal simplicity constraints on all of $H$ \footnote{To see this, divide the sphere into two charts. Choose a recouping scheme in which all the representations $\lambda_1, \ldots, \lambda_m$ from chart $1$ are coupled to a single representation, which in turn is coupled to the representations on chart $2$. Simplicity of the intermediate spin follows from $(L_{1}^{[IJ}+\ldots+L_{m}^{[IJ})(L_{1}^{KL]}+\ldots+L_{m}^{KL]})=0$, which is contained in both maximally commuting subsets.}. Thus, for spherical topology, the maximal amount of off-diagonal simplicity constraints can be imposed. 
This further restricts the horizon state to a simple SO$(D+1)$ intertwiner. Such an intertwiner has only simple representations in a given recoupling scheme as intermediate representations, see \cite{BTTV} for details.  

Since there is a 1 to 1 correspondence of simple SO$(D+1)$ intertwiners to SU$(2)$ intertwiners by mapping the recoupling labels for a given recoupling scheme onto each other, the problem of finding the dimension of the horizon Hilbert space is reduced to the $3+1$-dimensional case \cite{EngleBlackHoleEntropy}. The remaining step is to find an analogous restriction to a finite level of the Chern-Simons theory in $3+1$ dimensions. The finite value of the level translates into an upper bound on the spins puncturing the horizon and thus also on the ones in the recoupling scheme of the intertwiner. Such a bound can also be derived from considering that the area of the horizon slice $H$ is finite. 
The $(D-1)$-area operator $\hat A$ is constructed in the same way as in $3+1$ dimensions as $\sqrt{\text{``flux squared''}}$, see e.g. \cite{BTTIII} for details. It operates diagonally on an edge labelled with a simple SO$(D+1)$ representation. The eigenvalue is given by $8 \pi G \beta \sqrt{\lambda(\lambda+D-1)}$, $\lambda \in \mathbb{N}_0$.
The total area of a horizon slice punctured by $P$ edges is thus
\be
	A_{H} =  8 \pi G \beta \sum_{i = 1}^P \sqrt{\lambda_{i}(\lambda_{i}+D-1)} \text{.}
\ee
It follows that the sum of the puncture labels, and accordingly also all the recoupling labels, are bounded by
\be
	\sum_{i=1}^P \lambda_i < \sum_{i=1}^P \sqrt{\lambda_{i}(\lambda_{i}+D-1)} = \frac{A_H}{8 \pi G \beta } \leq k  \text{.} \label{eq:Bound2+1}
\ee
for $k = \lceil A_H / 8 \pi G \beta \rceil$. The definition of $k$ is meant to resemble the level of the Chern-Simons theory in \mbox{$3+1$} dimensions. Given this bound, the dimension of the intertwiner space at $H$ for given puncture labels $\lambda_i$ is \cite{EngleTheSU(2)Black}
\be
	N(A_H,P, k) = \frac{2}{k+2} \sum_{d=1}^{k+1} \sin^2\left( \frac{\pi d}{k+2} \right) \prod_{i=1}^P \frac{\sin \left( \frac{\pi d d_i}{k+2}\right)}{\sin \left( \frac{\pi d}{k+2} \right) } \text{,} \label{eq:DimHorizon}
\ee
where $d_i = \lambda_i + 1$, due to the 1 to 1 correspondence between SU$(2)$ and simple SO$(D+1)$ intertwiners. Note that this mapping is given by $\lambda = 2j$ \cite{BTTV}.

\section{Entropy calculation}

In order to compute the entropy of the isolated horizon, we first have to decide over which bulk degrees of freedom we want to trace. 
{In any case, as mentioned in the previous section, we will trace over gauge related bulk states. This selects the SO$(D+1)$-invariant subspace of the horizon Hilbert space due to the constancy of the smearing functions of the Gau{\ss} constraint, see also the discussion in \cite{KrasnovBlackHolesIn, DonnellyVacuumEntanglementAnd}. Thus, the relevant horizon states are gauge invariant by themselves. 
Up to this, there are two main proposals for tracing: the first one is to sum over all spin networks with fixed total horizon area $A_H$, thus having a variable number of punctures. This approach has been employed in the original calculation \cite{AshtekarQuantumGeometryOf} and extended to the SU$(2)$ approach \cite{EngleBlackHoleEntropyFrom} in \cite{KaulQuantumBlackHole, KaulLogarithmicCorrectionTo}. We can readily apply it and obtain an entropy proportional to $A_H / \beta$ at leading order. 
For $\lambda = 1$ as the lowest possible representation label, we obtain the logarithmic correction $-3/2 \, \text{Log}(A)$ found by Carlip in any dimension \cite{CarlipLogarithmicCorrectionsTo} and thus confirm this calculation from an independent point of view. 

On the other hand, one can argue that an observer outside of the black hole can in principle measure the spin network state, that is the graph (up to spatial diffeomorphisms) and the spin / intertwiner labels. Then, the number of punctures would be fixed.
A more recent proposal along these lines involving an analytic continuation of the Barbero-Immirzi parameter $\gamma$ to the (anti)-selfdual case $\pm i$ seems promising \cite{FroddenBlackHoleEntropy}. See \cite{BSTI, BNI, PranzettiBlackHoleEntropy} for more discussion on this route. The essential ingredients to this approach are to use a fixed number of punctures and the large spin limit. 
The method of \cite{FroddenBlackHoleEntropy} to analytically continue $\gamma$ in $3+1$ dimensions can be applied also in our dimension-independent treatment. In analogy to \cite{FroddenBlackHoleEntropy}, we continue $\beta$ to $\pm i/2$ and interpret $8 \pi G \sum_{i = 1}^P j_i$ as the area, since $j \gg 1$ \footnote{Note that the area spectrum for the SO$(4)$-based connection
  variables is double of that of the SU$(2)$-based variables, resulting from the fact that the representations in the SO$(4)$ theory are
  restricted to $\lambda/2 = j_+ = j_-$ \cite{BTTIII}.
}. Since the dimension of the horizon Hilbert space for given puncturing spins is the same as in the $3+1$-dimensional case, the results of \cite{FroddenBlackHoleEntropy} can be taken over verbatim, resulting in   
\be
 	S = \log N(A_H,P, \mp i \beta k/2)  = \frac{A_H}{4 G} + \text{corrections} \text{.}
\ee

\section{Polyhedral viewpoint}

A polyhedral viewpoint of the above calculation can be established in the following way: We recall that SU$(2)$ intertwiners in the $3+1$-dimensional theory can be seen as the quantization of convex polyhedra in three dimensions \cite{BianchiPolyhedraInLoop}.
Consider a polyhedron in $D$ spatial dimensions with $P$ $(D-1)$-faces. By the Minkowski theorem \cite{MinkowskiAllgemeineLehrsatzeUber}, it is in 1 to 1 correspondence (up to translations) with a set of weighted normals $\tilde{s}^i_n = A_n s^i_n$ satisfying $\sum_n \tilde s_n^i = 0$, where $A_n$ is the area of a the $n$-th polyhedral face and $s^i_n$ is a unit vector with $i = 1,\ldots,D$. 
Let the polyhedron now live in a $D+1$ dimensional Euclidean space, that is we consider $\tilde s^I$, $I = 0,\ldots,D$ and introduce an additional normal per face labelled $n^I_n$, so that $n^I_n \tilde s_n^J \delta_{IJ} = 0$. The natural symplectic structure is now given by $\{\tilde{s}^I_m, n^J_n \} =  \delta^{IJ} \delta_{mn}$ and is similar to \eqref{eq:BoundaryPoisson}. 
 
In order to still be able to apply the Minkowski reconstruction, we need to demand that $n^I_n = n^I$ for a fixed normal $n^I$. We thus introduce the bi-vectors $L_n^{IJ}$ per face together with the diagonal simplicity constraints $L_n^{[IJ} L_n^{KL]} = 0$ and off-diagonal simplicity constraints $L_n^{[IJ} L_m^{KL]} = 0$. These constraints ensure that $L_n^{IJ} = 2 n^{[I} \tilde s_n^{J]}$ \cite{FreidelBFDescriptionOf}. The symplectic structure of the bi-vectors is again similar to \eqref{eq:SympStrucL} and agrees with the one calculated from $n^I_n, ~\tilde s_n^J$ on the constraint surface of the simplicity constraints. We note the strong similarity to the symplectic structure in \cite{BianchiPolyhedraInLoop}. From the Minkowski reconstruction theorem, we have the additional constraint $\sum_n L_n^{IJ} = 0$.

As above, quantization of the Poisson-algebra of bi-vectors yields a Hilbert-space which is given by the product of individual SO$(D+1)$-representations. The quantization of $\sum_n L_n^{IJ} = 0$ projects this Hilbert-space onto its SO$(D+1)$-invariant part, while a maximally commuting subset of the simplicity constraints \cite{BTTV} yields a simple SO$(D+1)$-intertwiner, which is in 1 to 1 correspondence with an SU$(2)$-intertwiner. 
Thus, the foundation of the ideas discussed in \cite{BianchiPolyhedraInLoop} extends to higher dimensions. A thorough understanding this viewpoint may be key in better understanding the 1 to 1 mappings of the simple intertwiner spaces in higher dimensions found in \cite{BTTV}. 

As was discussed in \cite{BeetleGenericIsolatedHorizons, BeetleEntropyOfGeneric} in the four-dimensional context, the interpretation of the horizon states as different shapes of the horizon, i.e. polyhedra, seems in conflict with symmetry assumption made in the classical calculation. However, this tension was resolved in \cite{BeetleGenericIsolatedHorizons, BeetleEntropyOfGeneric} by showing that the classical calculation can be extended to arbitrary isolated horizons without changing the algebraic structure of the horizon degrees of freedom, thus allowing for the same quantization. It would certainly be interesting to extend this viewpoint to higher dimensions. In other words, one would like to show that the results of \cite{BTTXII, BNII} also hold for generic, e.g. rotating, isolated horizons, without changing the algebraic structure of the horizon degrees of freedom and the boundary condition. We leave this problem for further research.

\section{Comments}

	{\bf Lanczos-Lovelock gravity}, a generalization of higher-dimensional GR on the same phase space,  has been discussed in \cite{BNII} in the presence of an isolated horizon. The entropy calculation presented here is compatible with these results and the Wald entropy \cite{JacobsonEntropyOfLovelock, WaldBlackHoleEntropy} emerges at leading order  (with correct prefactor for the analytic continuation $\beta \rightarrow \pm i/2$). This is due to the fact that the area density $\sqrt{h}$ on $H$ is replaced by a Wald entropy density in the relevant calculations. 
	
	For {\bf connection variables} on the boundary, e.g. the Chern-Simons symplectic structures derived in \cite{BTTXII}, one could construct a locally gauge invariant off-diagonal horizon simplicity constraint by parallel transporting both $L^{IJ}$ to the same point. This way, local gauge invariance on $H$ might be kept. Though seemingly harder to achieve, a quantization along this route would be preferable. In contrast, the constraint used in this paper breaks local SO$(D+1)$-invariance on $H$ explicitly by dropping these parallel transporters. 
		
	The {\bf implementation of the simplicity constraints} proposed in \cite{BTTV} and used in this paper faces two potential problems in general situations: The chosen maximal subset might not commute with the action of the Hamiltonian constraint and the choice of subset might have an influence on the physics, i.e. different choices could be unitarily inequivalent at the level of observables. For the proposed application however, both objections do not apply, since the lapse function vanishes on $H$ and the dimension of the simple intertwiner space does not depend on the chosen subset. 
	
	In {\bf $2+1$ dimensions}, the simplicity constraints are automatically satisfied. If we still restrict to global gauge invariance on $H$ and apply the arguments of \cite{KrasnovBlackHolesIn}, the derivation goes through also here.

\section*{Acknowledgements}
\vspace{-2mm}
NB was supported by the NSF Grant PHY-1205388 and the Eberly research funds of The Pennsylvania State University. Discussions with Abhay Ashtekar, Marc Geiller, Alok Laddha and especially Yasha Neiman are gratefully acknowledged. Final improvements of this work were supported by a Feodor Lynen Research Fellowship of the Alexander von Humboldt-Foundation.

\end{document}